\documentclass[aps,prl,twocolumn,superscriptaddress,longbibliography,floatfix]{revtex4-2}
\usepackage{graphicx,bm,amssymb,amsmath,dcolumn,hyperref}
\usepackage{multirow}
\usepackage{enumerate}
\usepackage{enumitem}
\usepackage{color}
\usepackage{subfigure}  
\usepackage{epstopdf}
\usepackage{bbm}
\usepackage{graphicx}
\usepackage{threeparttable}
\usepackage{amsthm}
\usepackage{mathtools}
\usepackage{color}
\usepackage{tikz}
\usepackage{float}
\usepackage{empheq}

\def\eqa{\begin{eqnarray}}
	\def\eea{\end{eqnarray}}
\newcommand{\eq}{\begin{equation}}
	\newcommand{\ee}{\end{equation}}

\makeatother

\makeatletter

\begin{document}
	
	\title{Universal Scaling of Gap Dynamics in Percolation
	}
	
	\author{Sheng Fang}
	\thanks{These two authors contributed equally}
	\affiliation{School of Systems Science, Beijing Normal University, 100875 Beijing, China}
	
	\author{Qing Lin}
	\thanks{These two authors contributed equally}
	\affiliation{School of Systems Science, Beijing Normal University, 100875 Beijing, China}
	
	\author{Jun Meng}
	\affiliation{School of Science, Beijing University of Posts and Telecommunications, 100876 Beijing, China}
	
	\author{Bingsheng Chen}
	\affiliation{Network Science Institute and Department of Physics, Northeastern University, Boston, MA 02115,
		USA.}

	\author{Jan Nagler}
	\affiliation{Deep Dynamics, Centre for Human and Machine Intelligence, Frankfurt School of Finance and Management, Frankfurt am Main 60322, Germany}
	
	\author{Youjin Deng}
	\affiliation{Department of Modern Physics, University of Science and Technology of China, Hefei, Anhui 230026, China}
	\affiliation{Hefei National Laboratory, University of Science and Technology of China, Hefei 230088, China}

	\author{Jingfang Fan}
	\email{jingfang@bnu.edu.cn}
	\affiliation{School of Systems Science, Beijing Normal University, 100875 Beijing, China}
	\affiliation{Potsdam Institute for Climate Impact Research, 14412 Potsdam, Germany}
	
	\begin{abstract}    
		Percolation is a cornerstone concept in physics, providing crucial insights into critical phenomena and phase transitions. In this study, we adopt a kinetic perspective to reveal the scaling behaviors of higher-order gaps in the largest cluster across various percolation models, spanning from lattice-based to network systems, encompassing both continuous and discontinuous percolation. Our results uncover an inherent self-similarity in the dynamical process both for critical and supercritical phase, characterized by two independent Fisher exponents, respectively. Utilizing a scaling ansatz, we propose a novel scaling relation that links the discovered Fisher exponents with other known critical exponents.
		Additionally, we demonstrate the application of our theory to real systems, showing its practical utility in extracting the corresponding Fisher exponents. These findings enrich our understanding of percolation dynamics and highlight the robust and universal scaling laws that transcend individual models and extend to broader classes of complex systems.
	\end{abstract}
	\date{\today} 
	\maketitle


	{\it Introduction}.---
	Universality and scaling are two pillars in statistical physics,
	providing a framework for understanding the modern theory of critical phenomena~\cite{Stanley1999Scaling}. Percolation theory, as a prime candidate, has significantly contributed to our understanding of critical phenomena~\cite{stauffer2018introduction}, which describes the emergence of global connectivity with local links. 
	Through different rules, the percolation model has various extensions,  
	including explosive percolation (EP)~\cite{Achlioptas2009Explosive,da_costa_explosive_2010,nagler_continuous_2012,cho_avoiding_2013,DSouza2015Anomalous,DSouza2019Explosive,Li2023Explosive}, correlated percolation~\cite{Coniglio2009Correlated}, rigidity percolation~\cite{Jacobs1995Generic}, and directed percolation~\cite{Obukhov1980Problem} et al., even in high-order interaction networks~\cite{battiston_physics_2021}, which consider interactions among more than two nodes simultaneously, and multilayer networks, where multiple types of connections coexist~\cite{bianconi_percolation_2016, bianconi_multilayer_2018,de_arruda_fundamentals_2018}. 
	These models enrich the critical phenomena of percolation and have applications across a broad spectrum of modern science~\cite{Li2021Percolation,Fan2021Statistical,GoldenAckleyLytle1998,Davis2008,Boswell2008,SergeyStanley2010,GaoStanley2012,BashanHavlin2013,LiStanley2015,ZengStanley2020,ZhangStanley2019,BrunkZlotnick2018,BrunkTwarock2021,Herega2015,radicchi_percolation_2015,dong_resilience_2018,fan_climate_2018}, including material science, neuroscience, traffic, geophysics, epidemiology and more~\cite{cohen_complex_2010,saberi_recent_2015}. \par  
	
	For different percolation models, pursuing a unified framework to characterize its universal scaling behavior is challenging and extremely important. Recently, Fan \textit{et al.}~\cite{Fan2020Universal} provided a dynamic perspective on percolation by studying the scaling behavior of the largest gap, defined as the maximum increment in the size of the largest cluster $C(T)$ as bonds are gradually added over time $T$. They proposed three universal scaling functions to describe the critical probability distributions of the largest gap, showing their robustness across different percolation models. This raises fundamental questions: How robust is this framework when applied to higher-order dynamic gaps beyond just the largest one? Is there a new scaling law that governs the behavior of these higher-order gaps, and can it provide deeper insights into the critical dynamics of percolation systems?

	\begin{figure}[h]
		\centering
		\includegraphics[width=0.5\textwidth]{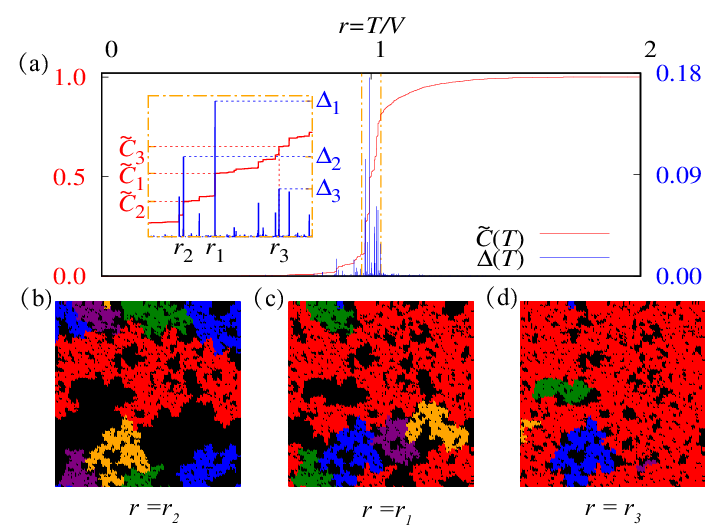}
		\caption{Illustration of the dynamical bond-adding process in bond percolation on a 2D square lattice with volume $V=128^2$. (a) The size of the normalized largest cluster $\tilde{C}(T) = C(T)/V$ is recorded dynamically, with its growth $\Delta(T)$ shown as the red and blue curves, respectively. Both curves share the same x-axis $r$, with the y1-axis (left) denoting the value of $\tilde{C}$ and the y2-axis (right) denoting the value of $\Delta$. The inset zooms in on the critical region, indicating the values of $\Delta_m$, $r_m$, and $C_m$ for $m \leq 3$. (b)-(d) Configurations at bond fractions $r=r_2$, $r_1$, and $r_3$, respectively, with the largest cluster highlighted in red.  
		}
		\label{fig1:env}
	\end{figure}
	
	In this Letter, we employ extensive simulations to investigate the scaling behaviors of higher-order gaps across a series of bond percolation models, extending beyond the focus on the largest gap. These models span a wide range, from lattice-based to network-based systems, and encompass both continuous and discontinuous percolation transitions: bond percolation on the square lattice (2D), bond percolation on the Erdős-Rényi graph (ER), EP with the product rule on the Erdős-Rényi graph~\cite{Achlioptas2009Explosive}, and restricted Erdős-Rényi graph (rER) with the fraction of restricted set $g=0.5$~\cite{Cho2016Hybrid}. In our simulations, as depicted in Fig.~\ref{fig1:env}(a), we dynamically record the size of the largest cluster $C(T)$ (red curve) and calculate the incremental change $\Delta(T) = \frac{G(T)}{V}= \frac{1}{V}[ C(T) - C(T-1) ]$ (blue curve) at each step $T$ with system volume $V$.
	For the $m$th-largest gap, we determine its location $r_m = T_m/V$ through the gap value $\Delta_m$, and subsequently calculate the corresponding largest cluster size $C_m = C(T_m)$. 
	The inset of Fig.~\ref{fig1:env} highlights the top three largest gaps, and Fig.~\ref{fig1:env}(b)-(d) display the configuration of 2D bond percolation for various $r$.
 \par 
	
	Through the numerical results, we find that for the location $r_m$ and the gap value $\Delta_m$, the scaling behaviors of their average values, fluctuations, and probability distributions are consistent for all gaps with finite $m$. 
	For the largest cluster $C_m$, although the average values and fluctuations are the same for all gaps, the probability distribution of $C_m$ ($m \geq 2$) exhibits a \textit{bimodal} distribution, which contrasts with the Gumbel distribution observed for the largest one, $C_1$. 
	Moreover, by analyzing the gap-size distribution $n(G,V)$, defined as the number density of gaps with size $G$, we uncover a self-similarity in the dynamic percolation process both at criticality and super criticality. This discovery leads to the proposal of two Fisher exponents and a modified hyperscaling relation, further enhancing our understanding of percolation dynamics. 
	In addition, these finding is further applied to percolation on real networks and investigate its critical behaviors.
	\par

	\begin{figure}[b]
		\centering
		\includegraphics[width=0.5\textwidth]{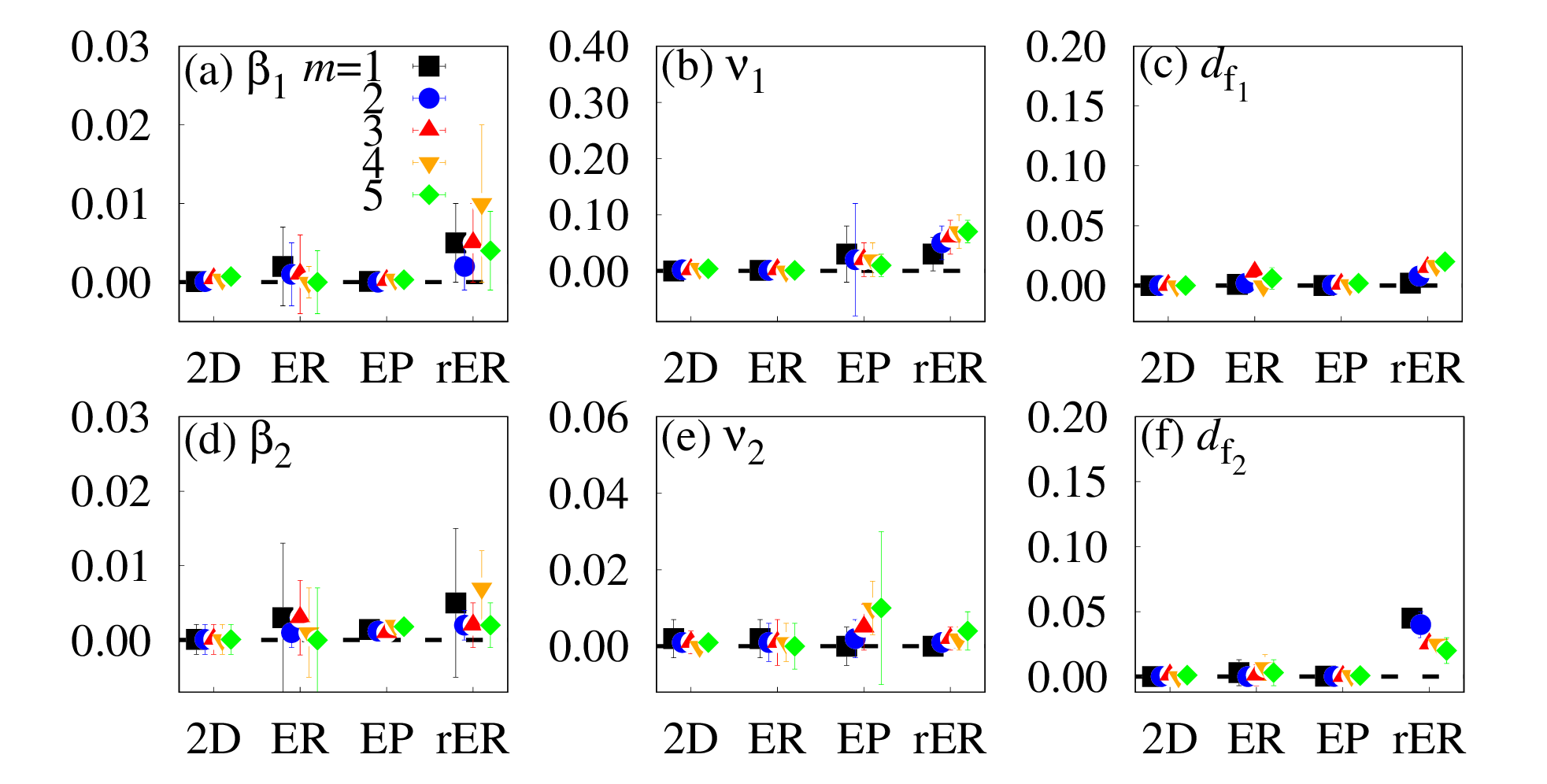}
		\caption{The absolute value of the deviation between the estimates of six critical exponents from theoretical prediction and previous estimates for 2D, ER, EP, and rER models. Detail estimates are shown in Tab.~\ref{tab:fit_expo}. }
		\label{fig:expoall}
	\end{figure}
	
	\par 
	
	{\it Scaling Behavior of Gaps}.---
	In the following, we consider the statistical properties, specifically the mean $\bar{(\cdot)}$ and fluctuation $\chi_{(\cdot)}$, of  $\Delta_{m}$, $r_{m}$ and ${C}_m$. For the largest gap ($m=1$), these quantities follow the scaling laws as reported in Ref.~\cite{Fan2020Universal}:
	\begin{align}
		\bar{\Delta}_1(V) \sim V^{-\beta_1}, \quad & \chi_{\Delta_1}(V) \sim V^{-\beta_2}, \nonumber \\ 
		|\bar{r}_1(V) -r_c| \sim V^{-1/\nu_1}, \quad & \chi_{r_1}(V) \sim V^{-1/\nu_2}, \nonumber \\ 
		\bar{C}_1(V) \sim V^{d_{f_1}}, \quad & \chi_{C_1}(V) \sim V^{d_{f_2}}.
		\label{eq:AveFlu}
	\end{align}
	Here, $r_c$ denotes the percolation threshold at the thermodynamic limit.
	The exponents $\nu_1 = \nu_2 = \nu$ represent the correlation length exponent, while the exponents $\beta_1 = \beta_2 = \beta/\nu$ represent the order parameter exponent divided by $\nu$, and $d_{f_1} = d_{f_2} = d_f$ represent the fractal dimension. 
	
	We then investigate the top five gaps of the largest cluster across all four models. Through a systematic finite-size scaling (FSS) analysis of simulation data, we obtain the estimates of the six exponents, which are summarized in Tab.~\ref{tab:fit_expo} in the Supplemental Material (SI)~\cite{Supplemental_Material}. 
	In Fig.~\ref{fig:expoall}, we present the deviation of all the exponents compared with their theoretical predictions or previous best estimates. We find that almost all of them are consistent with zero, indicating the robustness and universality of the scaling behavior for all gaps. The scaling laws governing the largest gap extend to higher-order gaps. 
	Notably, the exponents $\nu_1$, $d_{f_1}$, and $d_{f_2}$ of the rER model deviate slightly from the estimates, which may be due to unknown finite-size corrections, since it undergoes hybrid phase transitions~\cite{Cho2016Hybrid}. 	\par 
	
	\begin{figure}[b]
		\centering
		\includegraphics[width=0.5\textwidth]{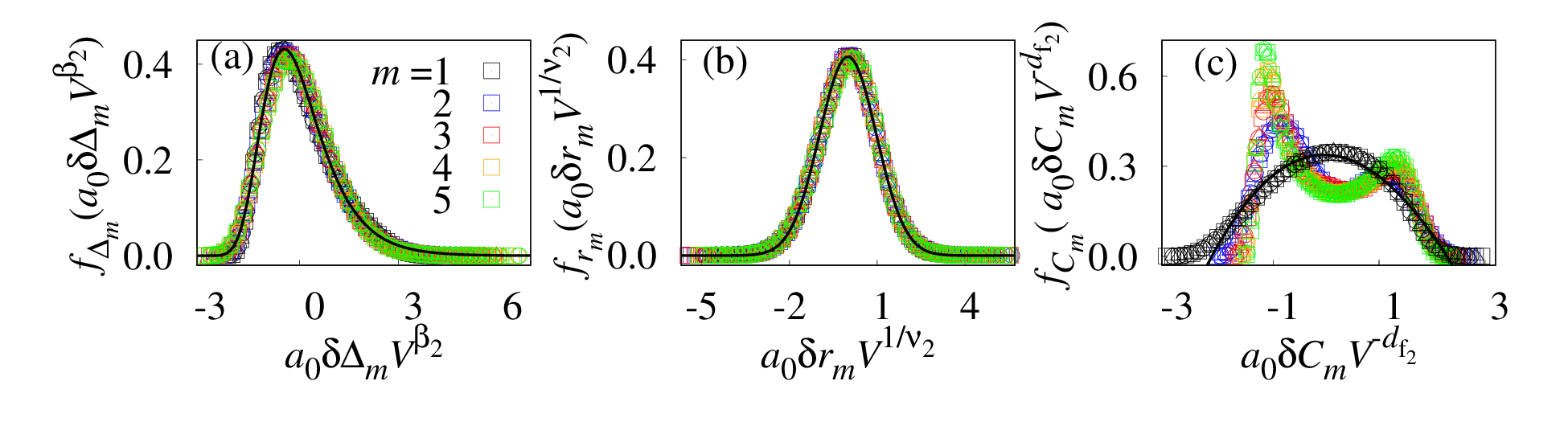}
		\caption{Universal scaling functions of (a) the gap value  $\Delta_m$, (b)  pseudo-critical point $r_m$, and (c) the largest cluster $C_m$ for the top $m$ gaps in bond percolation on a 2D lattice. Different orders $m$ are represented by different colors, and various system volumes are indicated by different point types. For each $m$, system volumes $V= 512^2$, $1024^2$, and $2048^2$ are used, denoted by $\square$, $\bigcirc$, and $\triangle$, respectively.The values of critical exponents, $\beta_2$, $\nu_2$ and $d_{f_2}$ are from Fig.~\ref{fig:2DAveFlu}. The $R^2$ values of fitting goodness are summarized in Tab.~\ref{tab:FitDis}. 
		}
		\label{fig:2DDisdelrS}
	\end{figure}

	{\it Universal gap scaling functions.}---
	We then consider the probability distributions of  $\Delta_m$, $r_m$, and $C_m$ for the top five gaps.
	These distributions are expected to take the following FSS forms~\cite{Fan2020Universal}:
	\begin{eqnarray}
		\label{Universal_1}
		P_{\Delta_m} (\delta\Delta_m, V) & = & V^{\beta_{2}}f_{\Delta_m}(\delta \Delta_m \cdot V^{\beta_2}),\\
		\label{Universal_2}
		P_{r_m}(\delta r_m,V) & = & V^{1/\nu_{2}}f_{r_m}(\delta r_m \cdot V^{1/\nu_2}),\\
		\label{Universal_3}
		P_{C_m}(\delta C_m,L) & = & V^{-d_{f2}}f_{C_m}(\delta C_m  \cdot V^{-d_{f2}}),
	\end{eqnarray}
	where $f_{\Delta_m}(\cdot)$, $f_{r_m}(\cdot)$ and $f_{C_m}(\cdot)$ are  universal scaling functions, and $\delta$ denotes the deviation from the average value. Specifically, $f_{r_1}$ follows the \textit{Gaussian} distribution, while the scaling functions $f_{\Delta_1}$ and $f_{C_1}$ approximately obey the \textit{Gumbel} distribution~\cite{Fan2020Universal,feshanjerdi_universality_2021}. 
	In Fig.~\ref{fig:2DDisdelrS}(a) and (b), we show the scaling functions  $f_{\Delta_m}$ and $f_{r_m}$ for the top five gaps on the 2D lattice. With an adjusted parameter $a_0$, we observe that data from various gaps ($m \in\{1,2,3,4,5\}$) and system volumes ($V \in \{512^2, 1024^2, 2048^2\}$) collapse well, and the scaling functions can be approximately described by the Gumbel and Gaussian distributions, respectively. (Deviations are observed in the tails of the curves, which is discussed in SI~\cite{Supplemental_Material} in detail.) 
	This demonstrates the robustness and universality of the scaling laws governing these distributions, Eqs.~(\ref{Universal_1})-(\ref{Universal_2}), extending the framework established for the largest gap to higher-order gaps. \par

	Renormalization group theory reveals that the size distribution of the largest cluster in subcritical percolation obeys the Gumbel extreme (or Fisher-Tippett-Gumbel) distribution~\cite{gyorgyi_finite-size_2008,sornette1998discrete}. Similarly, the probability distribution $f_{C_1}$ of the largest cluster $C_1$ at the largest gap $r_1$ also follows the Gumbel distribution approximately~\cite{Fan2020Universal}, as indicated by the black curve in Fig.~\ref{fig:2DDisdelrS}(c). For other gaps $m>2$, we find that our numerical data can also collapse onto a universal scaling function for various $V$. However, $f_{C_m}$ cannot be described by a Gumbel distribution (single peak) but instead exhibits a bimodal distribution  (see Fig.~\ref{fig:2DDisdelrS}(c)). 
	This bimodal phenomenon may arise from the increased variability of the pseudo-critical points $r_{m}$, as $\chi_{r_m}$ increases with $m$ (Fig.~\ref{fig:2DAveFlu}(e)), which causes higher fluctuations in $C_{m}$. 
	The bimodal distribution of the largest cluster has also been observed in the EP model~\cite{Li2023Explosive} and the geometric Ising model~\cite{Hou2019Geometric,Fang2021Percolation}. 
	
	The bimodal distribution reflects the competition between different dynamical processes governing cluster formation and growth. Near the pseudo-critical points, some clusters grow rapidly by merging with nearby clusters, while others grow more slowly or fragment. This dynamic interplay results in the observed bimodal size distribution, indicating that higher-order gaps capture a richer and more complex set of growth behaviors compared to the largest gap alone.
	To model this behavior, we introduce a generalized Gumbel distribution with two peaks, as described in Eq.~\eqref{eq:GGE}, and the goodness-of-fit is significant, as summarized in Tab.~\ref{tab:FitDis}.  
	Beyond the 2D lattice, we observe similar behaviors in other models, such as the ER, EP, and rER models~\cite{Supplemental_Material}. This indicates that the presence of a bimodal distribution for higher-order gaps is a robust and universal feature across different percolation systems, further extending the framework of universal gap scaling. 
	\par 
	
	\begin{figure*}[t]
		\centering
		\includegraphics[width=1.0\textwidth]{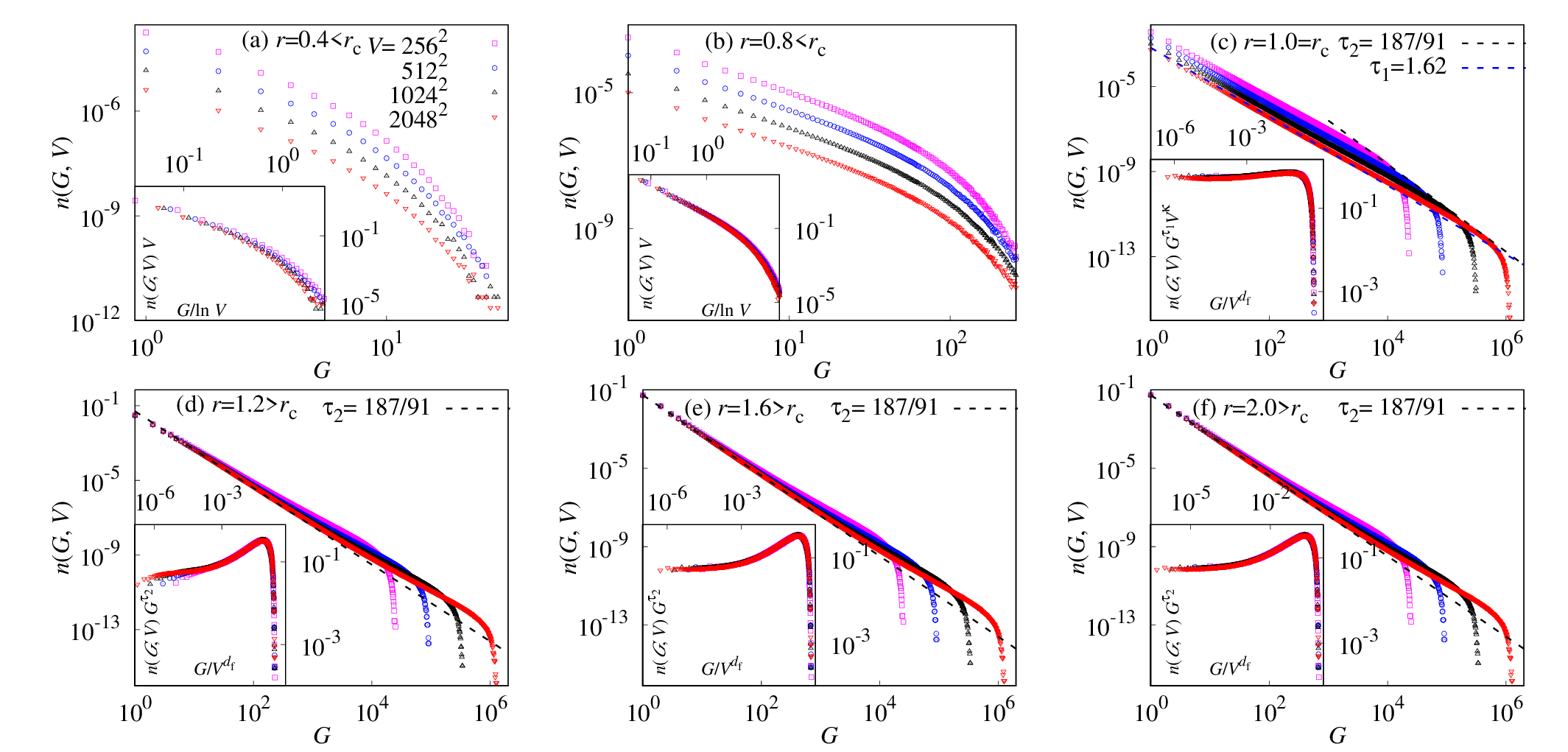}
		\caption{Gap-size distribution for bond percolation on a 2D lattice at different bond fractions $r =:$ (a) 0.4, (b) 0.8, (c) 1.0 (critical case), (d) 1.2, (e) 1.6 and (f) 2.0, respectively. The power-law exponent in (d)-(f), $\tau_2=187/91$ is consistent with the Fisher exponent of 2D critical percolation.
			The inset figures indicate that the gap-size distribution obeys Eq.~(\ref{eq:nGV}).  
		}
		\label{fig:2DGapSizeDis}
	\end{figure*}
	
	{\it Self-similarity for the percolation dynamic process}.--- We observe that the average values and locations of the top five gaps exhibit the same scaling behavior, i.e. $\Delta_m \sim V^{d_f} \sim C_m$ and $\delta r_m \sim V^{-\nu_1}$ ($1 \le m \le 5)$.  
	Then, we wonder how many gaps display the same scaling behaviors and whether they all fall within the critical window. 
	Before that, according to FSS theory near criticality, the distribution of finite components follows~\cite{stauffer2018introduction}, 
	\begin{equation}
		n(s,V) = n_0 s^{-\tau} \tilde{n}(s/V^{d_f}), 
		\label{eq:nsV}
	\end{equation}
	where $n_0$ is the amplitude, and the Fisher exponent $\tau$ is related to the fractal dimension via the hyperscaling relation,
	\begin{equation}
		\tau = 1+ 1/d_f.
		\label{eq:HyperScaling1}
	\end{equation}
	Thus, we obtain that the number of clusters of order $O(V^{d_f})$ remains constant. 
	This implies that the top $m$ clusters exhibit the same scaling behavior for finite $m$, indicating a \textit{self-similar} property.
	Compared with critical clusters, we note gaps may exhibit self-similar property in the dynamic process. \par

	We hypothesize that the scaling behavior of the top five gaps can be generalized to the top $m$ gaps for finite $m$. 
	Similar to the cluster-size distribution $n(s, V)$, we introduce the gap-size distribution $n(G, V)$ at the bond fraction $r$, which includes all the gaps of the largest cluster with bond fraction from $0$ to $r$.
	In Fig.~\ref{fig:2DGapSizeDis}(c), we present the gap-size distribution $n(G, V)$ at $r_c$, which displays a similar scaling behavior as $n(s, V)$. 
	For each system size, the distribution initially follows a power-law scaling with a finite cut-off. The cut-off is of order $O(V^{d_f})$, since $G_1 \sim V^{d_f}$. The power-law exponent $\tau_1$ is close to $1.62$, which is different from the 2D Fisher exponent $\tau = 187/91$ and indicates a breakdown of the hyperscaling relation in Eq.~(\ref{eq:HyperScaling1}).
	Moreover, the amplitude of $n(G,V)$ depends on the system volume $V$  rather than being constant. Similar to the procedure in Ref.~\cite{Hu2016NoEnclave}, we propose the gap-size distribution follows 
	\begin{equation}
		n(G,V) = n_0 V^{-\kappa} G^{-\tau_1} \tilde{n}(G/V^{d_f}).  
		\label{eq:nGV}
	\end{equation}
	When $\kappa=0$, this formula reduces to Eq.~(\ref{eq:nsV}).
	
	We further inquire about the values of the exponents $\kappa$ and $\tau_1$. Through scaling arguments~\cite{Supplemental_Material}, we posit that increments within the critical window (of order $O(V^{-1/\nu_1})$) dominate $n(G,V)$ for large $G$. Within the critical window, the total bond density scales as $V^{-1/\nu_1}$. Since the largest cluster scales as $C \sim V^{d_f}$ in this window, the probability that a randomly chosen bond belongs to $C$ is of order $O(V^{d_f-1})$. Thus, the gap-size distribution follows $n(G,V) \sim V^{-1/\nu_1 + d_f - 1}$, leading to $\kappa = 1/\nu_1 + 1 - d_f$.
	Based on this generalized assumption, we calculate that the number of gaps of order $O(V^{d_f})$ remains constant. Consequently, we modify the hyperscaling relation to
	\begin{equation}
		\tau_1 = 1 + \frac{1-\kappa}{d_f} = 2- \frac{1}{\nu_1 d_f} = 2-\sigma.   
		\label{eq:Hyperscaling2}
	\end{equation}
	where $\sigma = \frac{1}{\nu_1 d_f}$ is the critical exponent characterizing the divergence of the typical cluster size in the cluster number density $n(s)$. In 2D, this gives $\tau_1 = 2 - \sigma = 146/91 \approx 1.604\cdots$, which is consistent with our observations.
	To further refine the estimate for $\tau_1$, we compute the total number of gaps $N_G = V \sum_G n(G,V)$, which is expected to scale as $N_G \sim V^{1-\kappa}$, based on Eq.~\eqref{eq:nGV}. Our numerical results yield $N_G \sim V^{1-\kappa}$ with $\kappa = 0.41(1)$, resulting in $\tau_1 = 1.62(1)$. In the inset of Fig.~\ref{fig:2DGapSizeDis}(c), we plot $n(G,V) V^{\kappa} G^{\tau_1}$ versus $G/V^{d_f}$, and the data collapse confirms our estimate. Similar behaviors have been observed for $r=r_1$~\cite{Supplemental_Material}.
	\par  
	
	In addition, we investigate $n(G,V)$ for various bond fractions $r= $0.4, 0.8, 1.2, 1.6, and 2.0, as shown in Fig.~\ref{fig:2DGapSizeDis}. For $r<r_c$, the gap-size distribution $n(G,V)$ exhibits a power-law scaling with a cut-off size of order $O(\ln V)$, and the data collapse well when plotted as $n(G,V)V$ versus $G/\ln V$, as shown in the insets of Figs.~\ref{fig:2DGapSizeDis}(a) and (b), where the factor $1/V$ originates from the probability of the occurrence of gap in the subcritical phase.
	For $r>r_c$, the gap-size distribution $n(G, V)$ can be described by Eq.~\eqref{eq:nGV} with the exponent $\kappa=0$ and the power-law exponent $\tau_2 = 187/91$, consistent with the 2D Fisher exponent $\tau$, as indicated in  Figs.~\ref{fig:2DGapSizeDis}(d)-(f). 
	The alignment of the two exponents from the  hyperscaling relation Eq.~\eqref{eq:HyperScaling1} and $n(G,V)$ saturates for small $G$, i.e. $\kappa =0$, since $C$ spans the whole lattice and $N_G \sim V$ in the supercritical phase. These findings indicate the self-similarity occurs both at criticality and supercriticality. Moreover, the estimate of $\tau_2$ can be away from critical threshold.
	\par

	Besides the 2D lattice, we also investigate $n(G, V)$ for the ER, EP, and rER models.
	We observe similar scaling behaviors with estimated exponents summarized in Tab.~\ref{tab:NGV_exponent}. 
	For all four cases,  the exponents $\tau_2$ and $\tau_1$ fulfill the scaling relations Eqs.~\eqref{eq:HyperScaling1} and ~\eqref{eq:Hyperscaling2}, respectively, which confirms the hypothesis that the universal and robust scaling behaviors are beyond the top five gaps and further reinforcing the universality of scaling behavior across these models.
	\par

	\begin{table}[]
		\centering
		\begin{tabular}{|c|c|c|c|c|c|}
			\hline
			&  2D & ER & EP & rER  \\
			
			\hline 
			$\kappa$   & 0.41(1)   & 0.668(2) &  0.82(1)  & 0.855(1) \\
			\hline
			$\tau_1$  & 1.62(1)  & 1.498(4) &  1.19(1)  & 1.145(1)  \\
			2-$\sigma$ & 146/91  & 3/2      &  1.208(2)~\cite{Li2023Explosive} & 1.25(8)~\cite{Cho2016Hybrid}   \\
			\hline
			$d_f$   & 91/96     &  2/3     & 0.935(1)~\cite{Li2023Explosive} & 1.0(1)~\cite{Cho2016Hybrid} \\
			\hline 
			
			$\tau_2$  & 2.055(1)  &2.499(4)  &2.064(6)  & 2.18 \\
			& 187/91    & 5/2      &2.070~\cite{Li2023Explosive}  & 2.18~\cite{Cho2016Hybrid}   \\
			\hline
		\end{tabular}
		\caption{The critical exponents of the gap-size distribution.  The exponent $\kappa$ are estimated through the fitting results of $N_G$, where the former is consistent with previous results. The Fisher exponent $\tau_2$ and $\tau_1$ are obtained  through the hyperscaling relations Eq.~\eqref{eq:HyperScaling1} and Eq.~\eqref{eq:Hyperscaling2}, respectively. Note that for the rER model, the exponent $\tau_2$ and $d_f$ does not fulfill the hyperscaling relation Eq.~\eqref{eq:HyperScaling1}, since it presents the two scaling behavior, see Fig.~\ref{fig:rERBond_His}(c).}
		\label{tab:NGV_exponent}
	\end{table}

		Our framework is not only limited to theoretical percolation models but can also be applied effectively to real-world networks, such as social networks, collaboration networks, and scientific computing networks~\cite{Supplemental_Material}. Through this approach, we have successfully extracted the corresponding Fisher exponents, $\tau_1$ and $\tau_2$, demonstrating the practical utility of our theoretical model. This validation highlights the framework's relevance beyond purely abstract systems, making it a valuable tool for investigating critical phenomena in real-world complex systems, which is largely subjected to the estimation of critical point and failure of traditionary FSS procedure~\cite{Radicchi2015Percolation}. 
		By addressing this long-standing challenges, our work offers method for uncovering scaling behaviors and phase transitions in diverse, interconnected systems. \par

		\textit{Conclusion.}--- In this work, we generalize the kinetic framework  in \cite{Fan2020Universal} from the largest gap  to the finite top $m$ gaps across various percolation models, and find the scaling behaviors are universal and robust. Further, we uncover the self-similar properties of the dynamic percolation process through the gap-size distribution, with two independent Fisher exponents $\tau_2$, $\tau_1$ and hyperscaling relations $\tau_2 = 1+1/d_f$, $\tau_1 =2-\sigma$. 
		We emphasize that the estimates of two Fisher exponents are far beyond traditional procedure, which is free of the system volume and exact critical threshold. 
		Moreover, the independence of the two Fisher exponents indicates that all other common critical exponents can be obtained through the known scaling relations derived from renormalization group theory.  
		These superiority make it available to explore the critical phenomena in real systems, overcome the deficiency in first-order transition, and improve the determinations of critical exponents in real materials~\cite{Wang2020TwoScale}. 
		These findings offer deeper insights into the role of connectivity and dynamic transitions in shaping the critical behavior of gaps, and provide a new perspective and method on the critical phenomena and scaling laws. \par

	\section*{Acknowledgments}
	The authors would like to thank Prof. J\"urgen Kurths for his valuable discussions. J.F acknowledges the support of the National Natural Science Foundation of China (Grant No. 12275020, 12135003, 12205025) and the Fundamental Research Funds for the Central Universities. 

%
	
	\newpage
	\clearpage 
	\mbox{} 
	\setcounter{section}{0} 
	\setcounter{figure}{0} 
	\setcounter{equation}{0} 
	\setcounter{table}{0} 
	\renewcommand{\thefigure}{A\arabic{figure}}
	\renewcommand{\thetable}{A\arabic{table}}
	\renewcommand{\theequation}{A\arabic{equation}} 
	
	\onecolumngrid 
	\begin{center}
		\textbf{\large End Matter} 
	\end{center}
	~\\[1pt]
	
	\textit{Appendix A}. In this part, we display the conjectured scaling functions $f_{r_m} (\cdot)$, $f_{\Delta_m}(\cdot) $, and $f_{C_m}(\cdot) (1 \le m \le 5)$ as  
	\begin{equation}
		f_{\Delta_m}(\delta \Delta_m \times V^{\beta_2})  = A e^{-e^{-z} - z}, \text{with }  z = (\delta \Delta_m \times V^{\beta_2} - B)/\omega, 
		\label{eq:Gumbel}
	\end{equation}
	
	\begin{equation}
		f_{r_m} (\delta r_m \times V^{1/\nu_2})= A e^{-z^2}, \text{with } z = (\delta r_m \times V^{1/\nu_2})/\omega.
		\label{eq:Gauss}
	\end{equation}
	\begin{equation}
		f_{C_m}(\delta C_m \times  V^{-d_{f_2}})  =  A_{1}e^{-e^{-z_{1}}-z_{1}} + A_{2}e^{-e^{-z_{2}}-z_{2}}, \text{with } z_{1,2} = (\delta C_m \times V^{-d_{f_2}}-B_{1,2})/\omega_{1,2}   
		\label{eq:GGE} 
	\end{equation}
	Here, $A$, $B$, $\omega$, and $A_1$, $A_2$, $B_1$, $B_2$, $\omega_1$, $\omega_2$ are parameters. For the scaling function $f_{C_m}$ with $m=1$, the coefficient $A_1$ or $A_2$ is set to 0, and the scaling function returns to the Gumbel distribution, like Eq.~\eqref{eq:Gumbel}. In Tab.~\ref{tab:FitDis}, we display the fitting goodness of the top-five gaps via the above conjectured scaling functions. 
	
	\begin{table}[h]
		\centering
		\begin{tabular}{|c|c|c|c|c|c|}
			\hline 
			Model  &  2D   &  ER  & EP  & rER  \\
			\hline 
			$f_{\Delta_m}(\cdot)$ &$R^2 > 0.99$ & $R^2 > 0.98$ &  $R^2 > 0.99$   & $R^2 > 0.91$    \\
			
			\hline 
			$f_{r_m}(\cdot)$      &$R^2 > 0.99$ & $R^2 > 0.99$ & $R^2 > 0.99$    &  $R^2 > 0.99$     \\
			
			\hline 
			& $R^2>0.98$  & $R^2>0.99$ & $R^2>0.96$ & $R^2> 0.88$   \\
			& $R^2>0.97$  & $R^2>0.98$ & $R^2>0.97$ & $R^2>0.87$  \\
			$f_{C_m}(\cdot)$ & $R^2>0.91$  & $R^2>0.97$ & $R^2>0.98$ & $R^2>0.96$  \\
			& $R^2>0.87$  & $R^2>0.98$ & $R^2>0.89$ & $R^2>0.97$  \\
			& $R^2>0.92$  & $R^2>0.96$ & $R^2>0.93$ & $R^2>0.95$  \\
			\hline 
		\end{tabular}
		\caption{Goodness of fit and significance. The scaling function $f_{\Delta_m}(\cdot)$ is fitted through the Gumbel distribution Eq.~\eqref{eq:Gumbel}, and the scaling function $f_{r_m}(\cdot)$ is fitted through the Gaussian distribution Eq.~\eqref{eq:Gauss}. For the largest cluster $C_m$, when $m=1$, it is fitted through the Gumbel distribution by set $A_1=0$ in Eq.~\eqref{eq:GGE}, and when $m \ge 2$, it is fitted through the generalized Gumbel distribution Eq.~\eqref{eq:GGE}. }
		\label{tab:FitDis}
	\end{table}
	\par 
	\newpage 
	\onecolumngrid 
	\renewcommand{\thefigure}{S\arabic{figure}}
	\renewcommand{\thetable}{S\arabic{table}}
	\renewcommand{\theequation}{S\arabic{equation}}
	\mbox{} 
	\begin{center}
		\textbf{\large Supplemental Materials: \\ Universal Scaling of Gap Dynamics in Percolation } \\
		Fang \textit{et al}
	\end{center}
	
	In this supplemental material, we present the detailed numerical results of the bond percolation on the square lattice (2D), bond percolation on the Erdős-Rényi graph (ER), explosive percolation (EP) on Erdős-Rényi graph, and bond percolation on the restricted  Erdős-Rényi graph (rER). 
	\par 
	\section{Average value and function of gap scaling}
\begin{figure*}[b]
		\centering
		\includegraphics[width=1.0\textwidth]{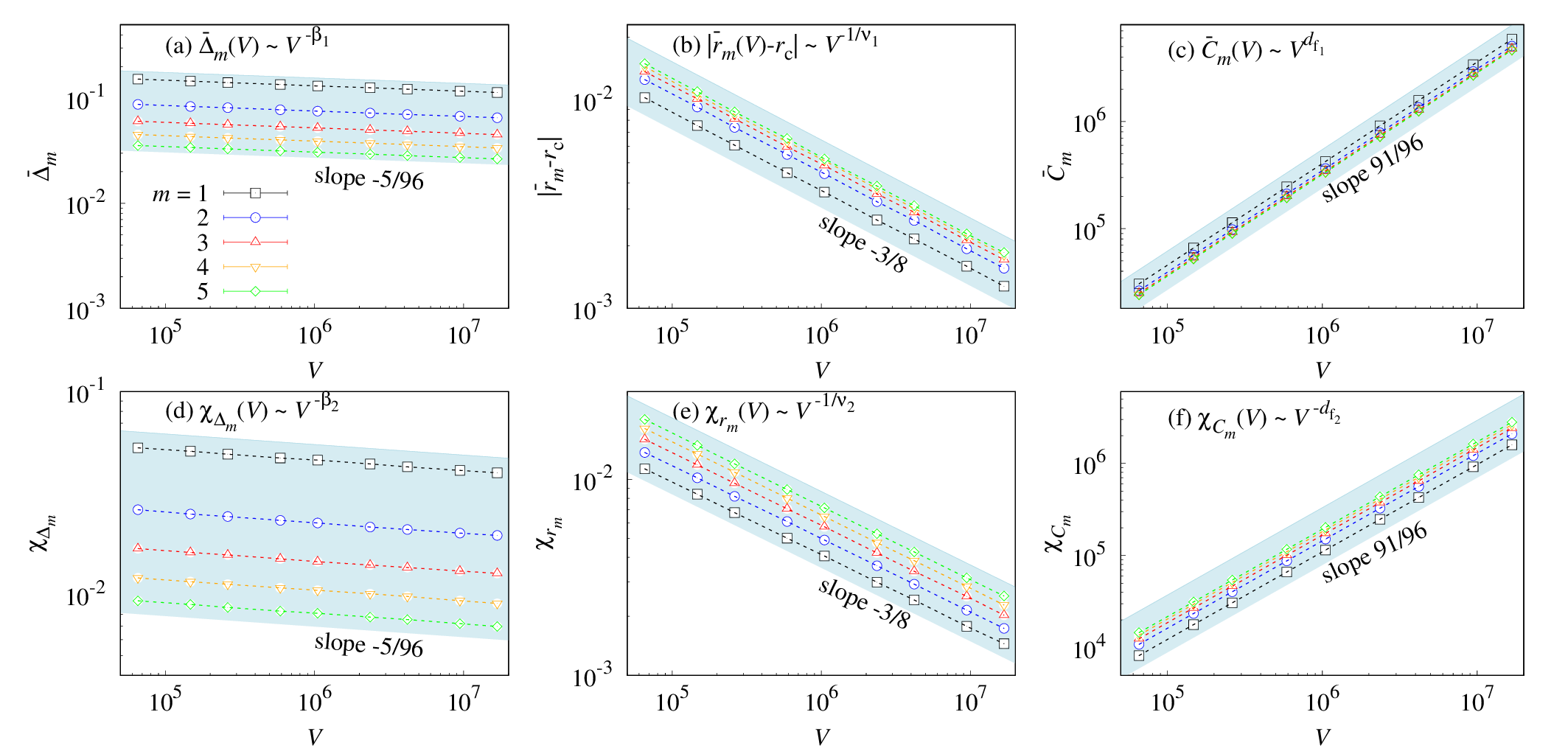}
		\caption{Average value and fluctuation of the top five gaps scaling for bond percolation on a 2D lattice with volume $V$ in log-log scale. (a)-(c)represent the average values of the gap $\bar{\Delta}_m$,  the pseudo-critical point $\bar{r}_m$ minus the percolation threshold $r_c$, and the size of the largest cluster associated with different gaps $C_m$, respectively. (d)-(f) illustrate their corresponding fluctuations. Error bars are too small to be visible in the plots. }
		\label{fig:2DAveFlu}
	\end{figure*}

\begin{figure*}
		\centering
		\includegraphics[width=1.0\textwidth]{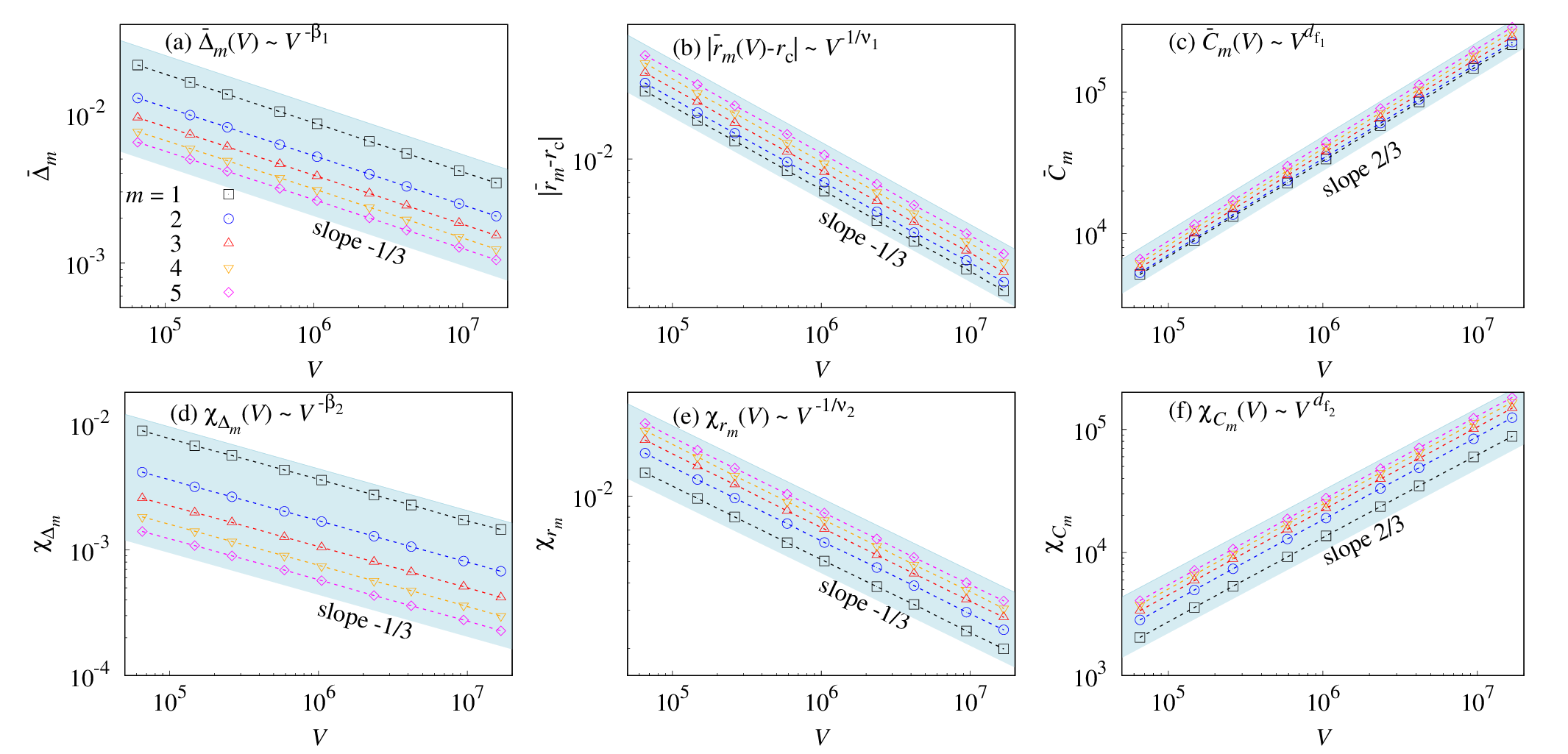}
		\caption{Average value and fluctuation of the top five gaps scaling for the ER bond percolation with volume $V$ in log-log scale. The graph setting is the same as Fig.~\ref{fig:2DAveFlu}.}
		\label{fig:ER_AveFlu}
	\end{figure*}

\begin{figure*}
		\centering
		\includegraphics[width=1.0\textwidth]{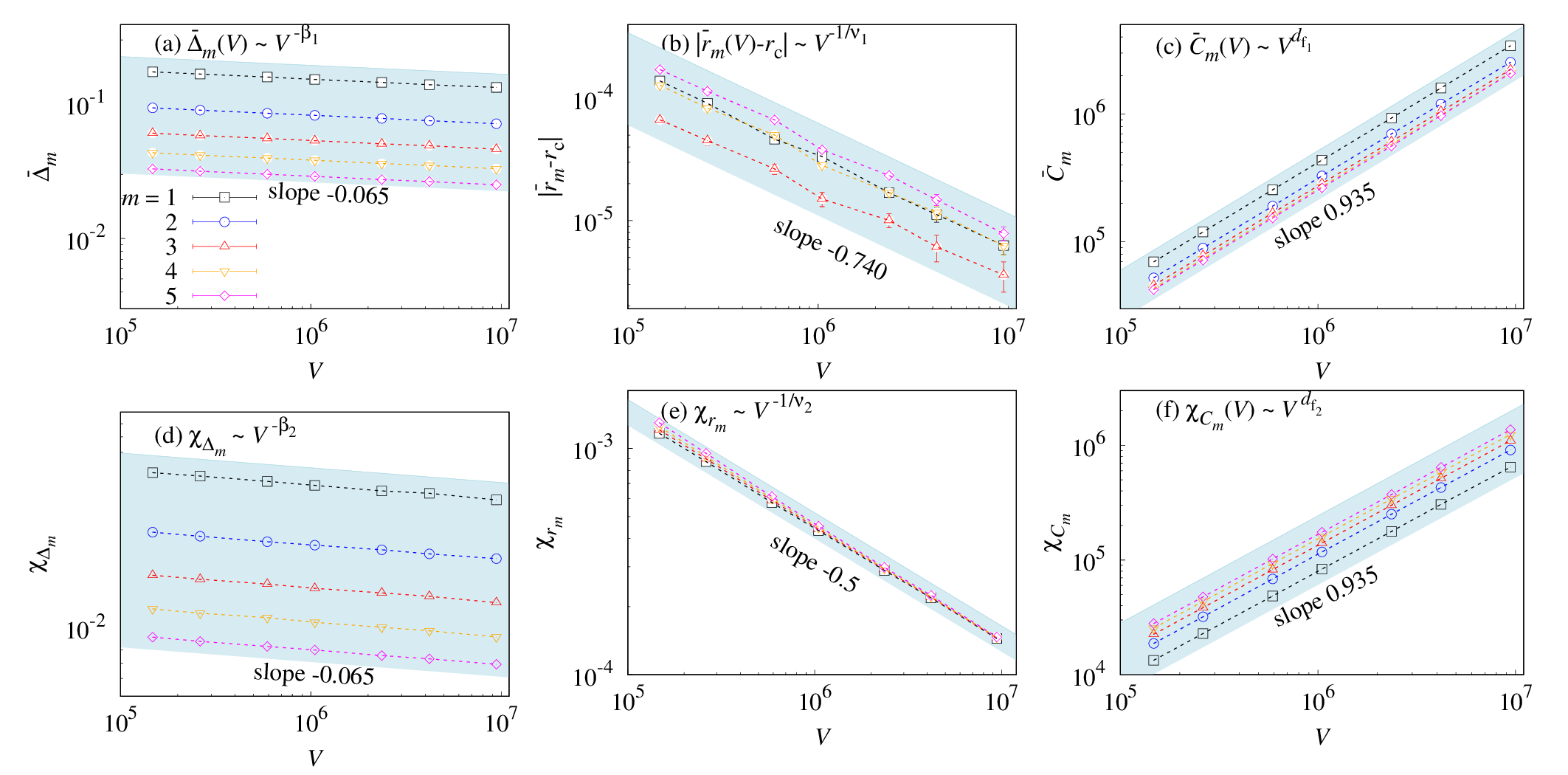}
		\caption{Average value and fluctuation of the top five gaps scaling for the EP percolation on the Erdős-Rényi graph versus volume $V$ in log-log scale. The graph setting is the same as Fig.~\ref{fig:2DAveFlu}. }
		\label{fig:EP_AveFlu}
	\end{figure*}

\begin{figure*}
		\centering
		\includegraphics[width=1.0\textwidth]{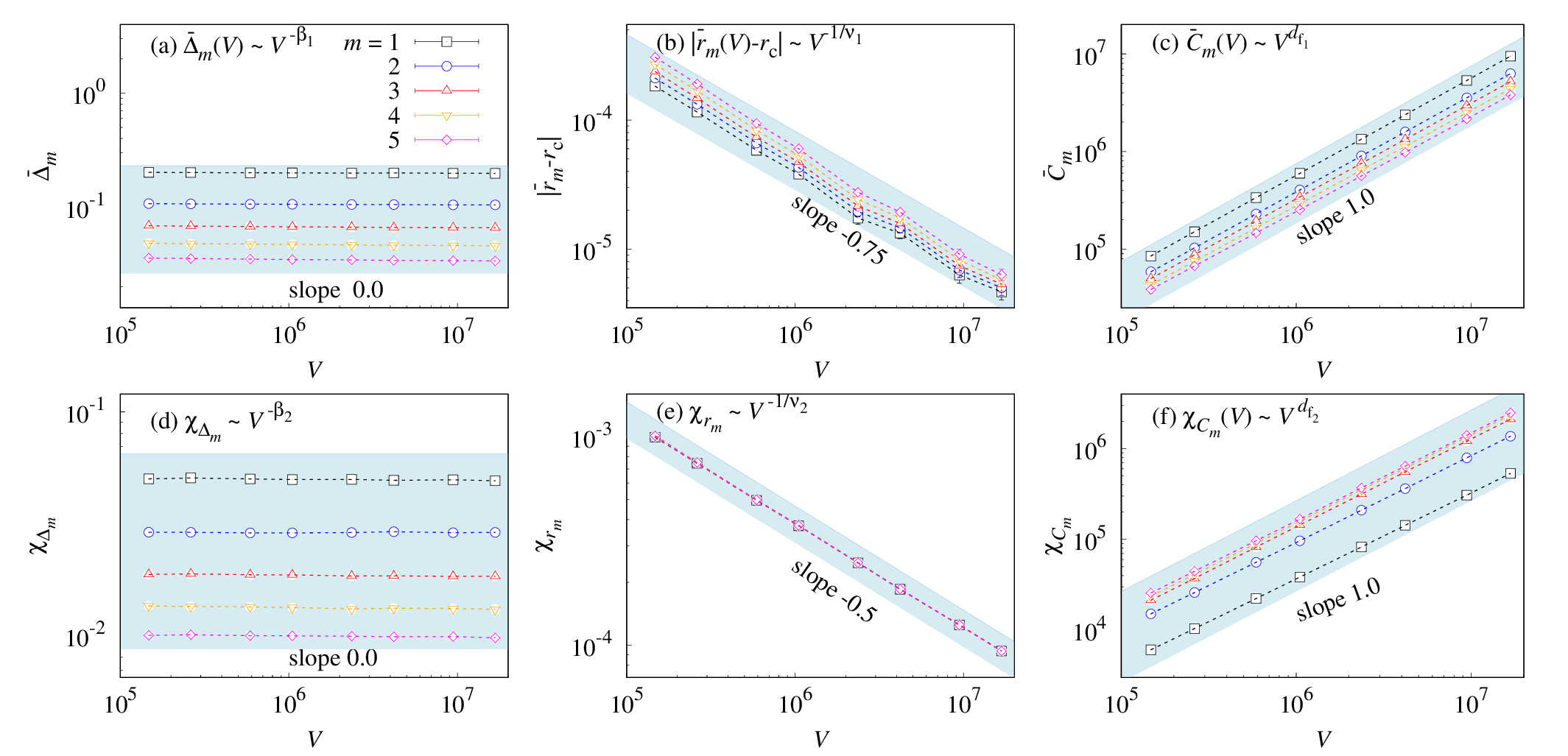}
		\caption{Average value and fluctuation of the top five gaps scaling for the rER model versus volume $V$ in log-log scale. The graph setting is the same as Fig.~\ref{fig:2DAveFlu}.}
		\label{fig:rER_AveFlu}
	\end{figure*}

	Figures~\ref{fig:2DAveFlu}-\ref{fig:rER_AveFlu} display the average value and fluctuation of the gap value $r_m$, gap location $\Delta_m$, and the largest cluster $C_m$  versus system volume $V$ in the log-log scale for the 2D, ER, EP, and rER models, respectively. 
	The subscript $m$ denotes the gap order. For each quantity in each studied system, different gaps are denoted by different point types. The light blue shadow draws the slope with the expected value from theoretical prediction and previous numerical estimates. 
	For all the four modes, the six exponents are almost consistent with previous results. We then further perform a least-square fit to the data and extract these exponents with reasonable errors through systematic analysis. The fitting results are summarized in Tab.~\ref{tab:fit_expo}. 
	For each model, the first line with bold font indicates theoretical prediction or previous estimate. 
	We find for 2D, ER, and EP models, all six critical exponents are consistent with previous results within two standard error bar. For the rER model, the exponents $\beta_1$, $\beta_2$, and $\nu_2$ are consistent with previous results, while for $1/\nu_1$, $d_{f1}$ and $d_{f2}$, it deviates from previous results, which may arise from some unknown corrections. 
	
	\begin{figure*} 
		\centering
		\includegraphics[width=\textwidth]{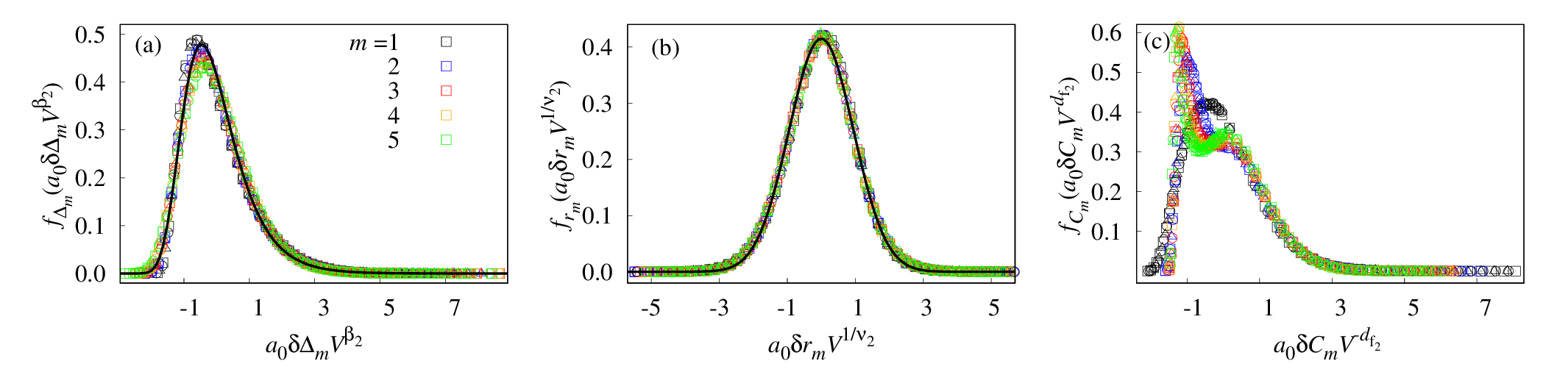}
		\caption{Universal scaling functions of (a) the gap value  $\Delta_m$, (b)  pseudo-critical point $r_m$, and (c) the largest cluster $C_m$ for the top $m (1 \le m \le 5)$ gaps in bond percolation on the ER graph. The graph setting is the same as Fig.~\ref{fig:2DDisdelrS}. The fitting goodness is summarized in Tab.~\ref{tab:FitDis}. }
		\label{fig:ER_Dis}
	\end{figure*}
	
	\begin{figure*}
		\centering
		\includegraphics[width=\textwidth]{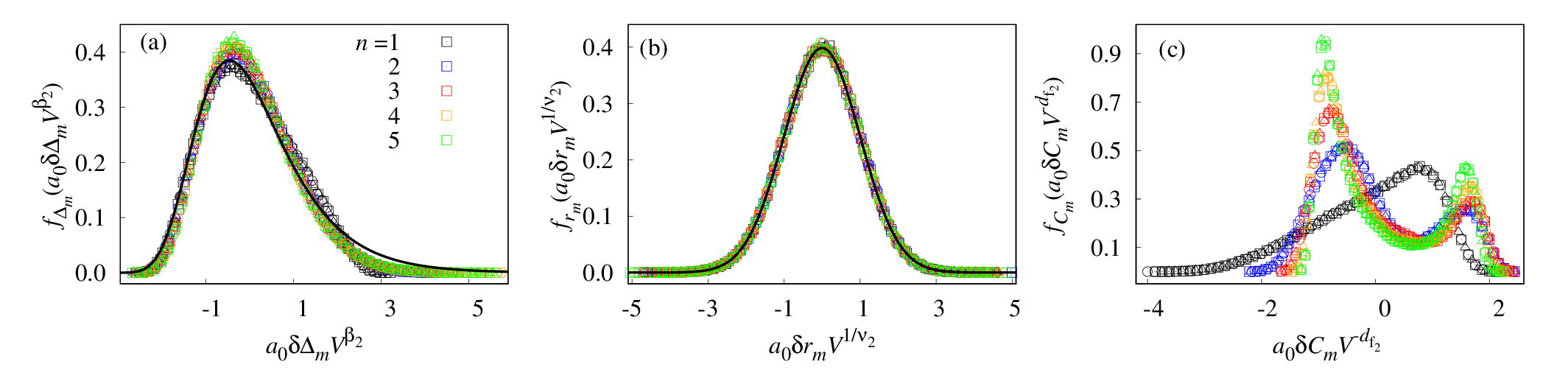}
		\caption{Universal scaling functions of (a) the gap value  $\Delta_m$, (b)  pseudo-critical point $r_m$, and (c) the largest cluster $C_m$ for the top $m (1 \le m \le 5)$ gaps in the EP model. The graph setting is the same as Fig.~\ref{fig:2DDisdelrS}. The fitting goodness are summarized in Tab.~\ref{tab:FitDis}.}
		\label{fig:EP_Dis}
	\end{figure*}

	\begin{figure*}
		\centering
		\includegraphics[width=1.0\textwidth]{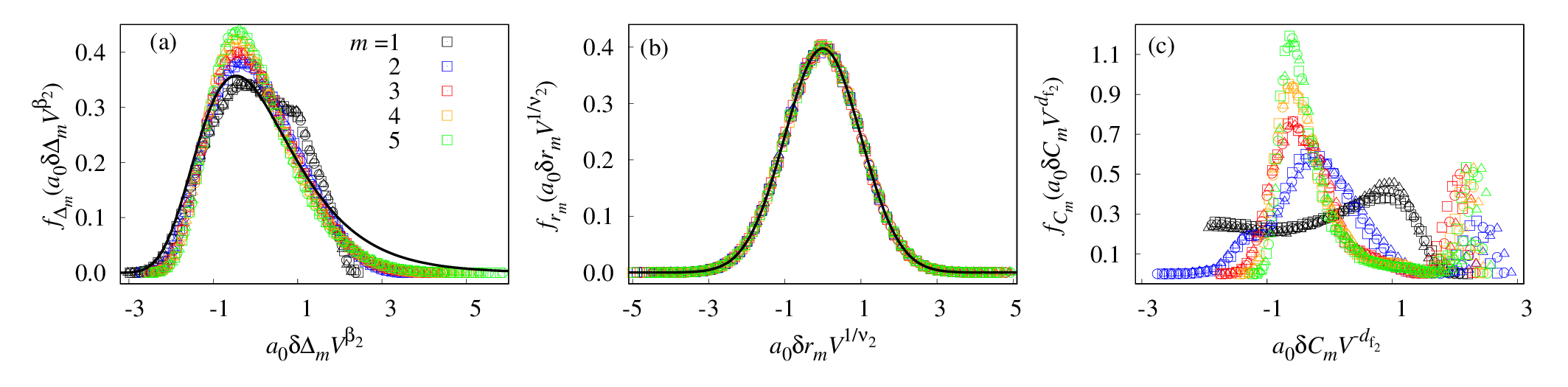}
		\caption{Universal scaling functions of (a) the gap value  $\Delta_m$, (b)  pseudo-critical point $r_m$, and (c) the largest cluster $C_m$ for the top $m (1 \le m \le 5)$ gaps in the rER model. The graph setting is the same as Fig.~\ref{fig:2DDisdelrS}. The fitting goodness are summarized in Tab.~\ref{tab:FitDis}.}
		\label{fig:rER_Dis}
	\end{figure*}
	
	\section{Probability distributions}
	
	In this section, we display the probability distribution of $\Delta_m$, $r_m$, and $C_m$. Figures~\ref{fig:ER_Dis}-\ref{fig:rER_Dis} display their distributions the ER, EP, and rER models, respectively. The setting of graphs is the same as Fig.~\ref{fig:2DDisdelrS} in the main text. 
	Different colors distinguish the top five gaps. For each gap, data from various system volumes with  $V= 512^2$, $1024^2$, and $2048^2$ can collapse well. With an adjusted parameter $a_0$, we can collapse the data of $\Delta_m$ and $r_m$ for all $1 \le m \le 5$. For the largest cluster $C_m$, they can not collapse well, even for $ m \ge 2$. \par  
	
	We assume the scaling function $f_{\Delta_m}(\cdot)$ and $f_{r_m}(\cdot)$ follow the Gumbel distribution and Gaussian distribution, as described in Eq.~\eqref{eq:Gumbel} and Eq.~\eqref{eq:Gauss}. 
	For the largest cluster $C_m$, it obeys the Gumbel distribution (Eq.~\eqref{eq:Gumbel}) for all the four models when $m =1$. While for $ m \ge 2$, it displays a binomial distribution. Through some conjecture, we assume it obeys a generalized Gumbel distribution as described in Eq.~\eqref{eq:GGE}. 
	Through the least-square fit to it, we obtain $R^2$ value of fitting results are summarized in Tab.~\ref{tab:FitDis}. 
	\par 
	
\begin{figure*}
		\centering
		\includegraphics[width=1.0\textwidth]{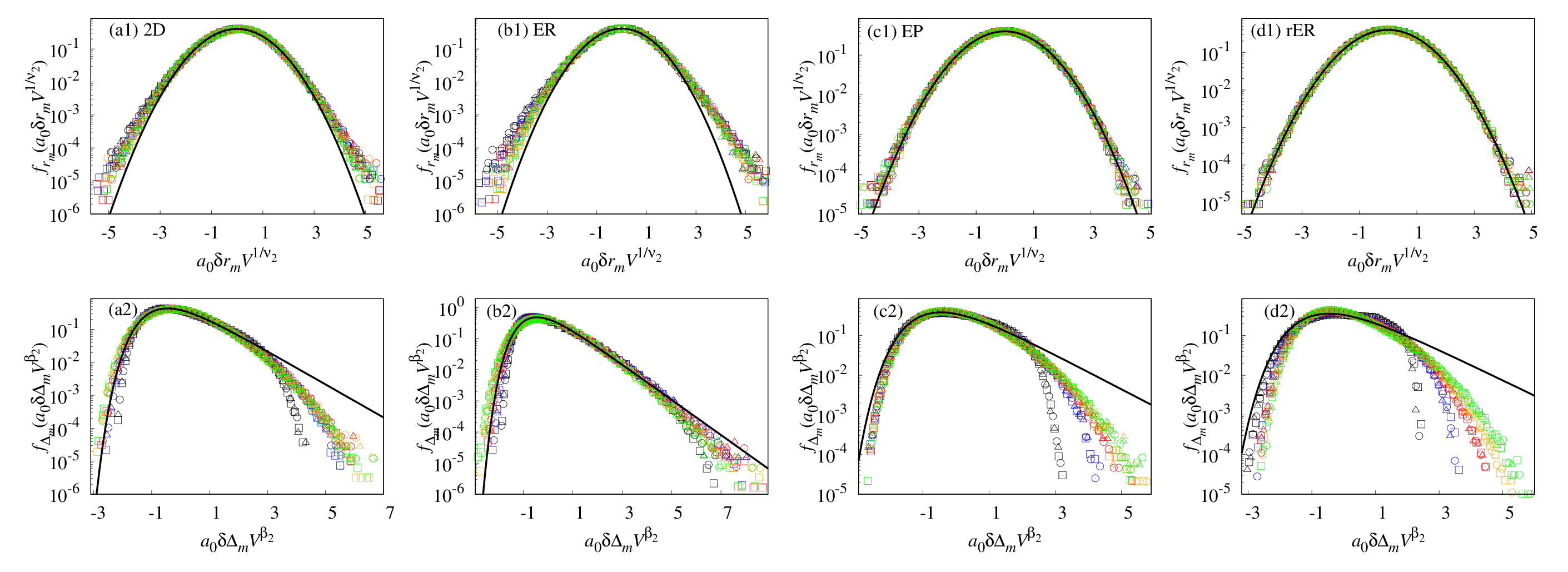}
		\caption{Universal scaling functions the pseudo-critical points $r_m$ for (a1) 2D, (b1) ER, (c1) EP, and (d1) rER models, respectively. Subfigures(a2)-(d2) present 
			the universal scaling functions of the gap value $\Delta_m$. Compared with Fig.~\ref{fig:2DDisdelrS} and Fig.~\ref{fig:ER_Dis}-Fig.~\ref{fig:rER_Dis}, the y axis is in the log scale, and one can see the deviations between data and fitting curves  are mainly from tails of the distributions.}
		\label{fig:Dislogy}
	\end{figure*}

	Here, we note that the probability distributions of gap location $r_m$, gap size $\Delta_m$, and the largest cluster $C_m$ approximately follow the proposed scaling functions, as indicated by the significant $R^2$ values in Tab.~\ref{tab:FitDis}. However, there are deviations of the data points from the fitting curves. In Fig.~\ref{fig:Dislogy}, we replot the probability distributions of $r_m$ and $\Delta_m$ for all four models with the y-axis on a logarithmic scale. We find that these deviations are primarily located in the tails of the distributions, where the values of the scaling functions are small. These deviations may originate from finite-size corrections. 
	Additionally, for the probability distribution of $r_m$ in the EP and rER models, the Gaussian distribution nearly covers the entire probability distribution, including the tail regions.
	
	\section{The gap-size distributions}
	In this section, we give the detail analysis of the gap-size distribution $n(G,V)$ for the ER, EP, and rER models.
	We first consider the conjectured formula $\kappa = 1/\nu_1 + 1-d_f$. 
	This conjecture is based on the assumption that the increment mainly comes from the bonds near critical point $r_c$, which is demonstrated as follows. 
	When the system is away from criticality, namely $t = r-r_c \sim$ O(1), the largest cluster is small and its increment is also small, which is bounded by O$(\rm{ln} V)$. The probability of a chosen bond belonging to the largest cluster is proportion to $1/V$, as shown in Figs.~\ref{fig:2DGapSizeDis}(a)(b). 
	As the bond density approaches the critical point at a slow rate $V^{-\lambda}$ with $\lambda < 1/\nu_1$, namely $r_c -r \sim V^{-\lambda}$. The largest cluster scale as $V^{\lambda \nu d_f}$ according to~\cite{Li2024Crossover}, thus, the gap-size density $n(G,V) \sim V^{-\lambda + \lambda \nu_1 d_f -1} = V^{-( \lambda + 1-\lambda \nu_1 d_f ) }$, where the exponent $( \lambda + 1-\lambda \nu_1 d_f )$ is larger than $\kappa$ for all $\lambda < 1/\nu_1$ and the increment outside the critical window contribute less to the gap-size distribution $n(G,V)$. 
	Additionally, We note other ways can also lead to the hyperscaling relation, as described by Eq.~\eqref{eq:Hyperscaling2}. For example, near criticality, one has $\bar{C} = V \sum_{G} G n(G,V) \sim V^{d_f}$, which can also leads to Eq.~\eqref{eq:Hyperscaling2}. 
	\par

		The dynamical process of gap formation and scaling can be described as follows: As the bond fraction increases, the largest cluster grows correspondingly. In the subcritical phase ($r<r_c$),  both the size and gap value of the largest cluster are bounded by  $O(\ln V)$, resulting in a limited number of gap events, with $N_G \sim O(1)$. This is because the probability that a newly added bond increases $C$ is of order $1/V$, meaning only a few events significantly affect the largest cluster's size at this stage. As $r$ approaches $r_c$, entering the critical phase, both the largest cluster size $C$ and the gap size $G$ increase sharply from $O(\ln V)$ to $V^{d_f}$. This phase transition causes a rapid escalation in the number of gap events, with $N_G$ increasing from $O(1)$ to $O(V^{1-\kappa})$. The sharp increase in cluster size reflects the system's progression toward criticality, where bond additions are increasingly likely to result in significant changes to the largest cluster. In the supercritical phase ($r > r_c$), the largest cluster spans the entire system, and the fraction of the system occupied by $C$ stabilizes. This implies that $C$ increases with almost every newly added bond, driving $N_G \sim V$. However, the incremental growth of $C$ becomes progressively smaller, preventing any further significant changes in the cut-off size of the gap distribution. As a result, the gap formation becomes less dynamic in this regime, and the scaling behavior stabilizes as the system saturates. 
	\par 
	
\begin{figure*} 
		\centering
		\includegraphics[width=\textwidth]{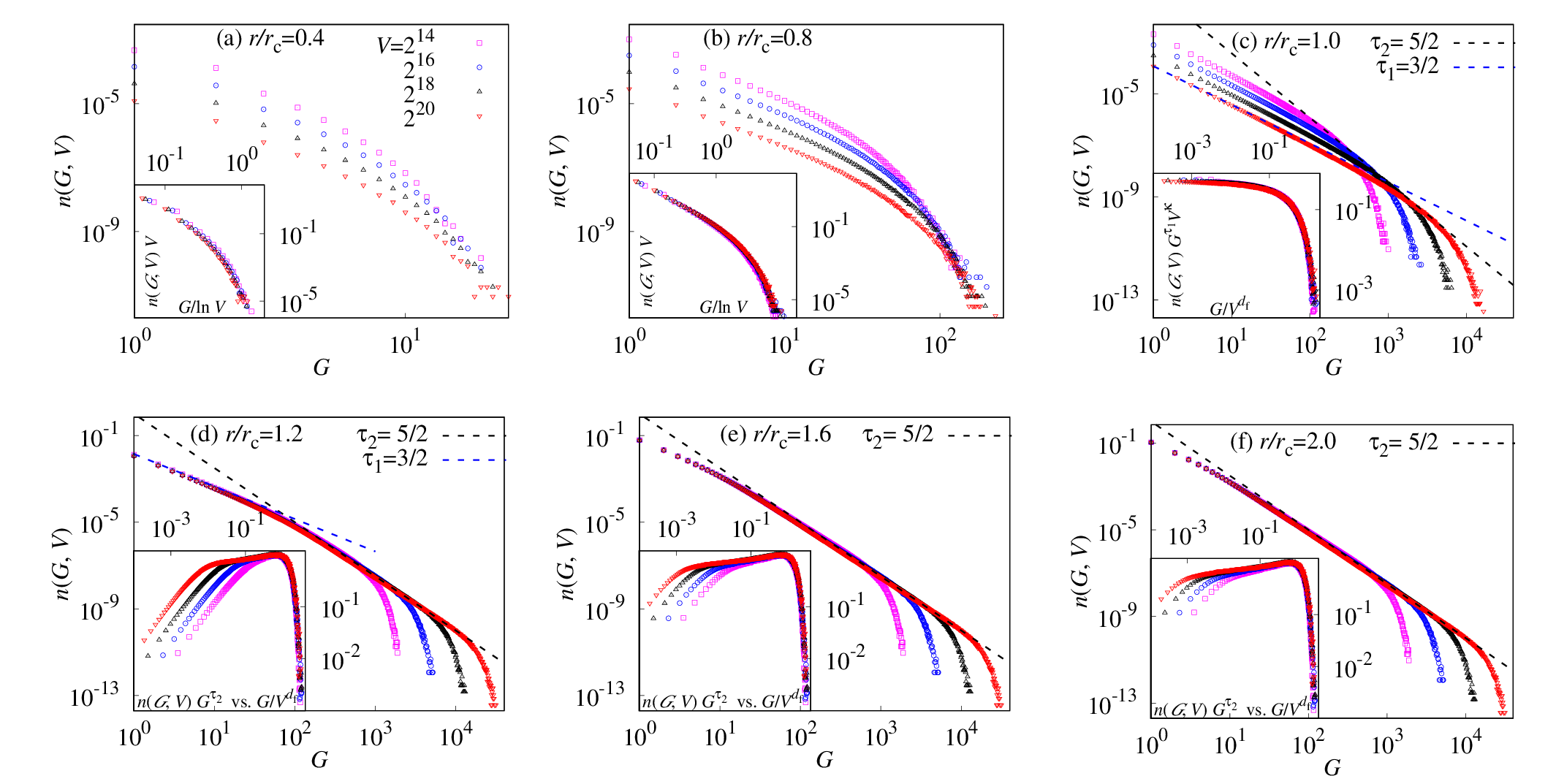}
		\caption{Gap-size distribution for bond percolation on the ER graph at different bond fractions $r/r_c =:$ (a) 0.4, (b) 0.8, (c) 1.0, (d) 1.2, (e) 1.6 and (f) 2.0, respectively. The power-law exponent in (d)-(f), $\tau_2=5/2$ is consistent with the Fisher exponent of ER critical percolation.
			The inset figures indicate that the gap-size distribution obeys Eq.~(\ref{eq:nGV}).}
		\label{fig:ERBond_His}
	\end{figure*}

\begin{figure*} 
		\centering
		\includegraphics[width=\textwidth]{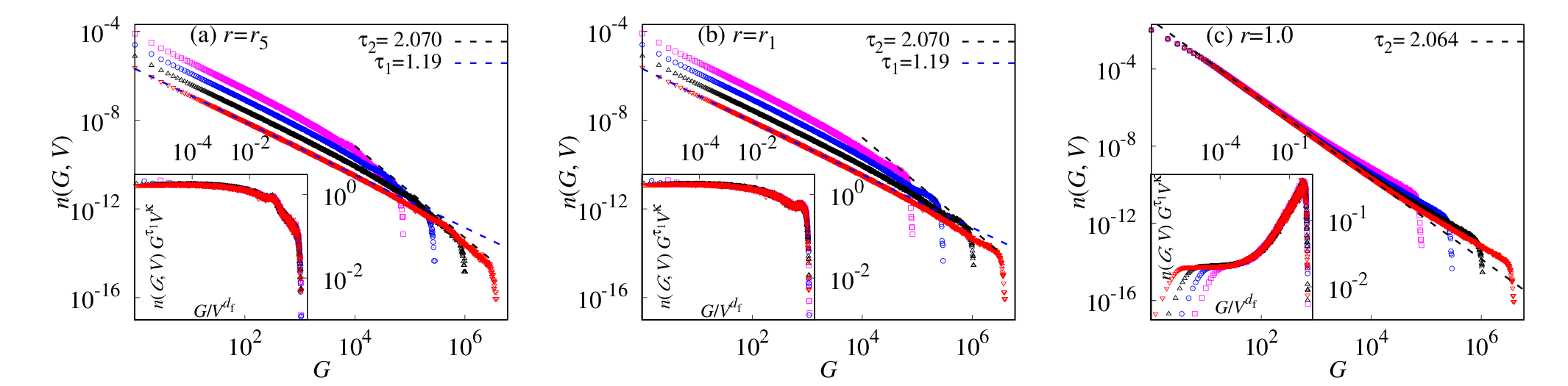}
		\caption{Gap-size distribution for EP model on the ER graph with the  bond fractions $r =:$ (a) $r_5$, (b) $r_1$, and (c) 1.0, respectively. The power-law exponent in (c), $\tau_2=2.064$ is consistent with the Fisher exponent of 2D critical percolation.
			The inset figures indicate that the gap-size distribution obeys Eq.~(\ref{eq:nGV}).}
		\label{fig:EPBond_His}
	\end{figure*}

\begin{figure*}
		\centering
		\includegraphics[width=1.0\textwidth]{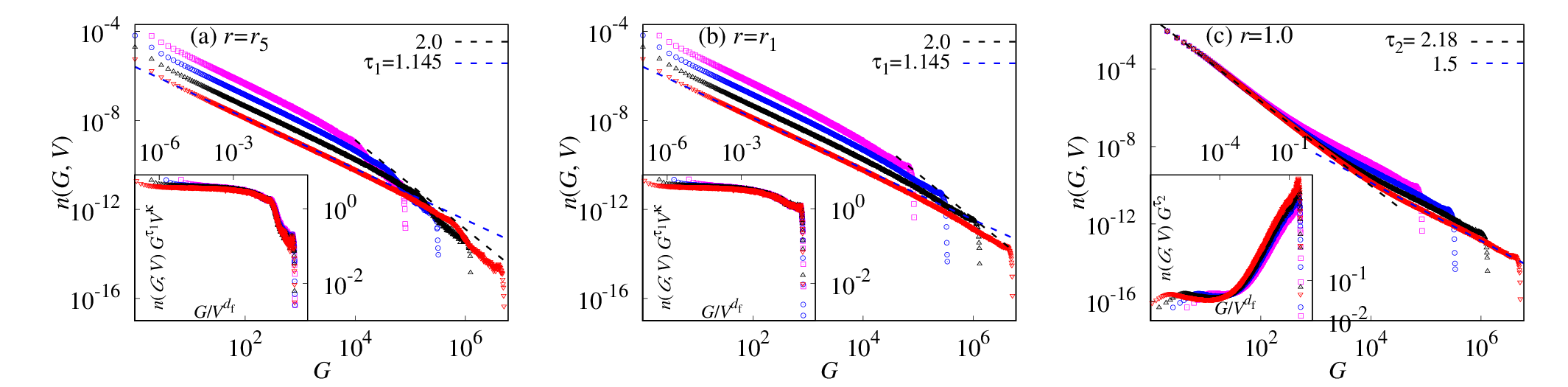}
		\caption{Gap-size distribution for rER model with the  bond fractions $r =:$ (a) $r_5$, (b) $r_1$, and (c) 1.0, respectively. The power-law exponent in (c), $\tau_2=-2.18$ is consistent with the Fisher exponent of rER model~\cite{Cho2016Hybrid}.
			The inset figures indicate that the gap-size distribution obeys Eq.~(\ref{eq:nGV}).}
		\label{fig:rERBond_His}
	\end{figure*}

	In Figs.~\ref{fig:ERBond_His}-~\ref{fig:rERBond_His}, we present the number results of the gap-size distribution $n(G,V)$ for ER, EP, and rER model, respectively, where similar scaling behaviors have be observed as those for bond percolation on 2D lattice. Following similar procedure, we obtain $\kappa = 0.668(2)$  for ER model. 
	While for the EP and rER models, we observe that $n(G,V)$ suffers larger correction at the threshold $r_c$ than $r_n$. Thus, in Fig.~\ref{fig:EPBond_His} and Fig.~\ref{fig:rERBond_His}, we display the $n(G,V)$ at $r=r_5$, $r=r_1$ and $r=1.0$ cases. Since $r_n$ is in the critical window, $n(G,V)$ presents the same scaling behavior with it at $r_c$. 
	Similarly, we obtain the estimate of exponent $\kappa = 0.82(1) $ and $0.855(5)$ through the FSS analysis of $N_G$. The good data collapse in Figs.~\ref{fig:EPBond_His}(a)(b) and Figs.~\ref{fig:rERBond_His}(a)(b) confirms our conjecture. 
	In addition, for the rER model, the $n(G,V)$ with $r=1.0$ presents a two scaling behavior, for $G$ is small, it decays with the power-law exponent $\tau_2$ consistent with 2.18, and when $G$  is large, the exponent is about $1.5$, as depicted in Fig.~\ref{fig:rERBond_His}(c), such that the data from various system size can not be collapsed well, as shown in the inset.

		Additionally, for the three models on the complete graph, the large number of edges significantly suppresses the growth of the largest cluster $C$, which results in a larger value of $\kappa$. This suppression occurs because the high connectivity in complete graphs limits the impact of adding new bonds on the size of $C$, reducing the number of large gaps that form. As a consequence, the number of gaps scales as $N_G \sim V^{1-\kappa}$, and since $\kappa$ is larger, the number of gaps $N_G$ is smaller compared to models with lower connectivity. In comparison with the ER model, the suppression of cluster growth is even more pronounced in both the EP and rER models due to the explosive nature of their phase transitions, leading to even larger values of $\kappa$. This suggests that high-connectivity networks and explosive percolation processes further constrain the formation of large gaps, reflecting a sharper distinction in their critical behavior.

	\begin{table}[]
		\centering
		\begin{tabular}{|c|l|l|l|l|l|l|}
			\hline 
			& ~~~~~$\beta_1$ & ~~~~~$\beta_2$ & ~~~~$1/\nu_1$ & ~~~~$1/\nu_2$ & ~~~~$d_{f_1}$  & ~~~~~$d_{f_2}$   \\
			\hline 
			& \textbf{5/96}  & \textbf{5/96} &\textbf{3/8}  &\textbf{3/8} & \textbf{91/96}  & \textbf{91/96} \\
			&    0.052\,0(5)  & 0.052(2) &0.375(5)   &0.373(5) &0.947\,9(2)  &0.948(2) \\
			&    0.052(1)  & 0.052(2) &0.377(3)      &0.374(2) &0.948\,0(4)  &0.948(2) \\
			2D &    0.052\,5(4) & 0.052(2) &0.377(3) &0.374(3) &0.948\,0(5)  &0.949(2) \\
			&    0.052\,5(5)  & 0.052(2) &0.370(5)  &0.375(1) &0.947\,9(4)  &0.948(1) \\
			&    0.052\,8(4) & 0.052(2) &0.379(4)   &0.374(1) &0.948(1)   &0.949(1) \\
			\hline 
			& \textbf{1/3}  & \textbf{1/3} &\textbf{1/3}  &\textbf{1/3} & \textbf{2/3}  & \textbf{2/3} \\  
			&    0.331(5) &  0.33(1) & 0.334(3)  &0.335(5)  &0.666(4) & 0.67(1) \\
			&    0.332(4) &  0.334(2) & 0.334(2)  &0.332(5)  &0.665(6) & 0.667(4) \\
			ER &    0.334(5) &  0.336(5) & 0.335(3)  &0.334(6)  &0.656(6) & 0.668(8) \\
			&    0.333(2) &  0.334(6) & 0.333(3)  &0.334(5)  &0.667(8) & 0.66(1) \\
			&    0.333(4) &  0.333(7) & 0.334(3)  &0.333(6)  &0.661(9) & 0.67(1) \\
			\hline 
			& \textbf{0.065(1)}  & \textbf{0.064(1)} &\textbf{0.75(1)}  &\textbf{0.500(1)} & \textbf{0.935(1)}  & \textbf{0.935(1)} \\  
			& 0.064\,9(5) & 0.065\,4(7)  & 0.72(5) &0.500(5)  & 0.9349(4) &0.935\,3(8) \\
			& 0.065\,0(3) & 0.065\,2(7)  &0.73(10) &0.498(5)  & 0.9345(5) &0.934\,8(7)  \\
			EP   & 0.065\,2(5) & 0.065\,0(9)  &0.77(3)  &0.495(6)  & 0.934(1)  &0.935\,0(9) \\
			& 0.065\,4(5) & 0.066(1)     &0.77(3)  &0.490(7)  & 0.934(1)  &0.934(1) \\
			& 0.065\,3(8) & 0.065\,8(8)  &0.76(2)  &0.49(2)   & 0.937(3)  &0.934\,2(8)  \\ 
			\hline
			& \textbf{0.0}  & \textbf{0.0} &\textbf{0.75(1)}  &\textbf{0.50(1)} & \textbf{1.0(1)}  & \textbf{1.0(1)} \\  
			&   0.005(5)  &  0.005(10) &  0.78(3) &  0.500(2) & 0.998\,0(5) & 0.955(3)\\
			&   0.002(3)  &  0.002(2)  &  0.80(3) &  0.499(2) & 0.992(3) &    0.96(1)\\
			rER  &   0.005(5)  &  0.002(3)  &  0.81(3) &  0.498(3) & 0.985(3) &    0.975(5)\\
			&   0.01(1)   &  0.007(5)  &  0.82(3) &  0.498(3) & 0.983(5) &    0.975(6)\\
			&   0.004(5)  &  0.002(3)  &  0.82(2) &  0.496(5) & 0.980(5) &    0.98(1)\\
			\hline
		\end{tabular}
		\caption{Fitting results of the six critical exponents for the 2D, ER, EP, and rER models. In each model, the first line with bold font originates from theoretical prediction for 2D and ER, and previous numerical estimate for EP~\cite{Fan2020Universal} and rER~\cite{Cho2016Hybrid}.}
		\label{tab:fit_expo}
	\end{table}
	
	\section{Application}

\begin{figure*}
		\centering
		\includegraphics[width=1.0\textwidth]{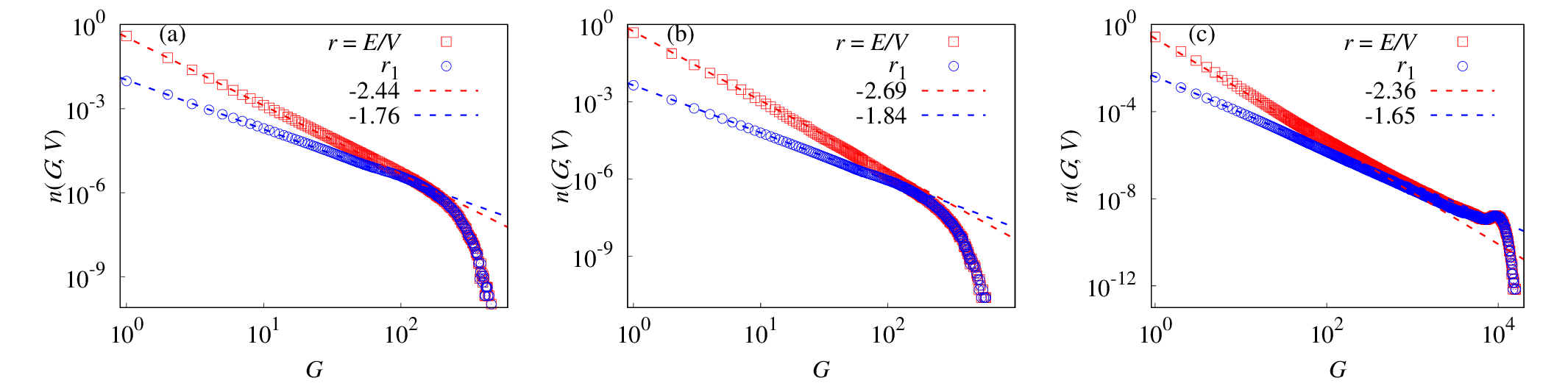} 
		\caption{Gap-size distribution for bond percolation with fraction $r= 1.0$ (red) and $r=r_1$ (blue) on three real networks~\cite{Ryan2015NetworkData}. (a) social networks with vertexes and edges $(V, E)= (3\,484, 155\,040)$; (b) Collaboration network of arXiv astrophysicists with $(V,E) = (17\,903, 196\,969)$; (c) scientific computing networks  with $(V,E)=(87\,804 ,2\,565\,054)$.} 
		\label{fig:NsRN}
	\end{figure*}

		In this section, we study the bond percolation on real networks and investigate the gap-size distribution $n(G,V)$. In  Fig~\ref{fig:NsRN}, we choose three different types of real networks~\cite{Ryan2015NetworkData}, including social networks, collaboration networks, and scientific computing networks. We measure the gap-size distribution $n(G,V)$ at the bond fraction $r=r_1$ and $r=E/V$ with total vortexes number $V$ and edges number $E$. For all three networks, the power-law exponents $\tau_1$ and $\tau_2$ are different. We then perform a least-square fit to the data, we then obtain the estimate $(\tau_1, \tau_2) = (1.76(5), 2.44(1))$, $(1.84(2), 2.69(5))$, $(1.65(1), 2.36(4))$, respectively. Through the hyperscaling relations in Eqs.~\eqref{eq:HyperScaling1} and~\eqref{eq:Hyperscaling2}, we obtain the critical exponents $d_f$ and $\nu$. Thus, all other critical exponents can be obtained through sophisticated scaling relations. We note this procedure provides a new perspective to describe network structure. 
	
\end{document}